# Manipulating Spin-polarized Photocurrent in 2D Transition Metal Dichalcogenides

Lu Xie and Xiaodong Cui[*]

Physics department, University of Hong Kong, Hong Kong, China

* To whom correspondence should be addressed.  Email: xdcui@hku.hk

**Abstract**: Manipulating spin polarization of electrons in nonmagnetic semiconductors by means of electric fields or optical fields is an essential theme of the conceptual nonmagnetic semiconductor-based spintronics. Here we experimentally demonstrate a method of generating spin polarization in monolayer transition metal dichalcogenides (TMD) by the circularly polarized optical pumping. The fully spin-polarized photocurrent is achieved through the valley dependent optical selection rules and the spin-valley locking in monolayer $WS_2$, and electrically detected by a lateral spin-valve structure with ferromagnetic contacts. The demonstrated long spin lifetime, the unique valley contrasted physics and the spin-valley locking make monolayer $WS_2$ an unprecedented candidate for semiconductor based spintronics.

A longtime focus in nonmagnetic semiconductor spintronics research is to explore methods to generate and manipulate spin of electrons by means of electric fields or optical fields instead of magnetic fields, enabling scalable and integrated devices.[1] The present efforts follow two distinct paths. One utilizes spin Hall effect or optical pumping in III-V semiconductors which feature a significant spin-orbit coupling in a form of Dresselhauls and/or Rashba terms;[2-4] The other focuses on spin transport (usually generated by spin injection from ferromagnetic (FM) electrodes) in semiconductor structures made of silicon[5], carbon nanotube[6], graphene[7], *etc* which have long spin coherence length due to weak spin-orbit coupling. The emergence of atomic 2 dimensional group VI transition metal dichalcogenides (TMDs) $MX_2$ (M=Mo, W; X=S, Se), featuring nonzero but contrasting Berry curvatures at inequivalent $K$ and $K'$(equivalent to $-K$) valleys and unique spin-valley locking, provides an alternative pathway towards spintronics.[8]

Valleys refer to the energy extremes around the high symmetry points of the Brillouin zone, either a "valley" in conduction band or a "hill" in valence band. Owing to their hexagonal lattices, the family of TMDs have degenerate but inequivalent $K(K')$ valleys well separated in the first Brillouin zone. This gives electrons extra valley degree of freedom, in addition to charge and spin. In monolayer TMDs the inversion symmetry breaking of crystal structures gives rise to nonzero but contrasting Berry curvatures at $K$ and $K'$ valley which are a characteristic of the Bloch bands and could be recognized as a form of orbital magnetic momentum of Bloch electrons.[9-11] These contrasting Berry curvatures of electrons (holes) at $K(K')$ valleys lead to contrasting response to certain stimulus.[9-19] One example is valley Hall effect: An electric field would drive the electrons at different valleys ($K$ and $K'$) towards opposite transverse directions, in a similar way as in spin Hall effect.[10,20] A more pronounced manifestation is valley-dependent circular optical selection rules in $K$ and $K'$ valleys. Namely the interband optical transitions at $K(K')$ only couple with circularly polarized light of σ+(σ-) helicity. Consequently the valley polarization could be realized by the polarization field of optical excitations.[10-13] On the other side, the band edge at $K(K')$ valleys mainly constructed from $d$ orbits of the heavy metal atoms inherits the strong spin-orbit coupling (SOC) of atomic orbits. And the Zeeman like SOC originating from $D_{3h}^1$ symmetry of monolayer TMDs lifts the out-of-plane spin degeneracy of the band edges at $K$ and $K'$ valleys by a significant amount, around 0.16eV and 0.45eV in the valence bands of molybdenum dichalcogenides and tungsten dichalcogenides, respectively, and about one order of magnitude smaller in conduction bands.[10,21-27] Owing to the presence of time reversal symmetry ($K \leftrightarrow -K$), the spin splitting has opposite sign between $K$ and $K'$ valleys at monolayer TMDs as illustrated in Figure 1a. The Kramer doublet, spin-up state $S_z = \frac{\hbar}{2}$ at $K$ valley $|K \uparrow\rangle$ and spin-down state $S_z = -\frac{\hbar}{2}$ at $K'$ valley $|K' \downarrow\rangle$, are separated from the other doublet $|K \downarrow\rangle$ and $|K' \uparrow\rangle$ by the SOC energy. This strong SOC and the explicit inversion symmetry breaking lock the spin and valley degrees of freedom in monolayer TMDs and this interplay leads to sophisticated consequences. First the spin and valley relaxation are dramatically squelched due to the simultaneously requirements of spin flip and momentum conservation. The intrinsic mirror symmetry with respect to out-of-plane direction further suppresses spin relaxation via D'yakonov-Perel' mechanism which usually plays an important role for spin relaxation in III-V semiconductors. Subsequently the valley and spin polarization are expected to be robust against low energy perturbation.[10,28] Second the spin-valley locking offers a versatile measure to manipulate spin degree of freedom via control of valley degree of freedom or vice versa.[8,29-31] This could lead to an integrated and complementary approach of valleytronics and spintronics in monolayer TMDs.

Here we report an experimental demonstration of spin polarization via valley-dependent optical selection rules in monolayer $WS_2$. The valley polarization is realized by controlling the polarization field of interband optical excitations and the spin polarization is simultaneously generated via spin-valley locking in monolayer $WS_2$. The spin polarization is electrically detected by the lateral spin-valve structure consisting of a tunneling barrier of $Al_2O_3$ and superlattice-structured Cobalt-Palladium (Co/Pd) ferromagnetic electrodes with perpendicular magnetization anisotropy (PMA). A near-unit spin polarization of the diffusive photocurrent is observed and a micron size spin free path and spin lifetime in range of $10^1 \sim 10^2$ ns are estimated.

The photocurrent measurements were conducted on a 5μm-channel field effect transistor (FET) structure of mechanically exfoliated monolayer $WS_2$. To overcome the conductance mismatch for efficient spin filtering, an ultrathin $Al_2O_3$ (1.2nm) was deposited between the monolayer and ferromagnetic electrodes made of Co/Pd superlattice. Owing to the intrinsic mirror symmetry ($D_{3h}^1$) with respect to the plane of metal atoms, the spin projection is along out-of-plane direction and $S_Z$ is a good quantum number in monolayer $WS_2$. To electrically detect the spin polarization along $z$ direction, a spin analyzer with PMA is the key, which is realized with a superlattice of ultrathin Co(4.5Å)/Pd(15Å) multilayers.[32] The *in-situ* polar magneto-optic Kerr effect spectroscopy (MOKE) demonstrates a clear ferro-magnetization along $\vec{z}$ direction with a coercive force around 30Oe, as shown in Figure 2b.(supplementary information) Standard electric characterization as shown in Figure 2a shows a slightly n-type FET behavior in all the devices, which might be induced from the defects, vacancy and/or substrate effects. The source-drain conductance is tens nano-siemens level at maximum within the back gate bias range of $V_G = \pm 80V$, showing the Fermi level falls deep in the band gap.

The drain current rises by 1-3 orders of magnitude when the near-resonance excitation scans across the monolayer, similar to the reported photocurrent experiments on multilayer $WS_2$.[33] As demonstrated in Figure 3b, the scanning photocurrent distributes inhomogeneously across the channel, concentrating around charge traps/defects and electrode contacts where local electric fields are strong enough to break excitons, quasiparticles of Coulomb-bounded electron-hole pairs, into free carriers. To generate significant photocurrents with a minimum background electric current (dark current), a gate $V_G = -80V$ pulls the FET to "off" state and a source-drain bias $V_{DS} = -5V$ is applied to accelerate the photo-carriers. Once the FM electrodes are ferro-magnetized by the external magnetic field, the photocurrent shows a distinct pattern of optical-polarization responses at zero magnetic field as demonstrated in figure 3c-d. For the excitation close to the electrode-TMD contacts, the strength of the photocurrent exhibits a strong dependence on the combination of the FM electrode magnetization and the polarization of optical excitations. Depending on the magnetization of FM electrodes, the photocurrent at the same location shows a clear circular dichroism for the circularly polarized optical excitations with opposite helicities. Namely under one magnetization direction, for example, along positive $\vec{z}$ ($M \uparrow$), the excitation with polarization of $\sigma^+$ induces higher photocurrent than that of $\sigma^-$. If the FM magnetization is reversed, the photocurrent difference ($\sigma^+ - \sigma^-$) between opposite helicities also switches the sign. The non-zero photocurrent difference shows a clear dependence on the magnetization of FM electrodes as shown in Figure 4b, which is consistent with the magnetization of the FM electrodes demonstrated in the MOKE measurements. The photocurrent difference has a clear spatial distribution pattern: it generally rises upon the excitation spot being close to FM electrodes and it vanishes when the excitation is far away from the FM electrodes. This scenario is well understood with the valley-dependent circular optical selection rules and the spin-valley locking in monolayer $WS_2$. The excitation of $\sigma^+(\sigma^-)$ selectively pumps the excitons at $K(K')$ valley and the electrons/holes are fully spin polarized to $S_z = \frac{\hbar}{2}$ ($S_z = -\frac{\hbar}{2}$) due to spin-valley locking. If local electric fields breaks the excitons into free carriers, these free carriers are accelerated by the source-drain bias to generate photocurrents while the spin polarization remains. If the spin polarization survives when the photo-carriers reach the FM electrodes, the spin alignment with the FM electrodes yields the different effective resistance. Figure 3a-d also show the photocurrents and the photocurrent difference $I_{\sigma+} - I_{\sigma-}$ are uncorrelated. It is because the scanning photocurrent directly reflects the strength of local electric fields, whereas the photocurrent difference $I_{\sigma+} - I_{\sigma-}$ also depends on the photocarriers' spin polarization arriving at the FM electrodes and the efficiency of the electrode-TMD junction for spin filtering.

To quantitative evaluate the spin polarization of the photocurrent, we define the degree of the photocurrent polarization $P = (I_{\sigma+} - I_{\sigma-})/(I_{\sigma+} + I_{\sigma-})$ where $I_{\sigma+}(I_{\sigma-})$ is the photocurrent under the excitation of $\sigma^+(\sigma^-)$. Given that the photocurrent $I_{\sigma+}(I_{\sigma-})$ at minimum is around nano-ampere which is far beyond the noise level of tens pico-ampere in the system, artifacts in calculating polarization P are safely excluded. The photocurrent polarization $P$ peaks around 0.16 at electrode-TMD junctions and decays to a negligible level when the optical excitation scans away from the FM electrodes as shown in Figure 3e, 3f and 4c. The polarization $P$ reverses the sign if the magnetization of FM electrodes switches, showing a signature of efficient spin-valve structure. As a result of valley-dependent optical selection rules and spin-valley locking in monolayer TMD, the photocurrent polarization $P$ reflects the spin polarization of the electrons (holes) arriving at the electrode-TMD junction. At the experimental conditions the photocurrent is dominated by the diffusive drift current (supplementary information), and the electrons/holes' trajectory could be simplified as a collective movement with a uniform velocity. Without considering many-body interactions, the spin polarization exponentially decays with a characteristic time, equivalently distributing with a characteristic spin polarization free path in space. Consequently the profile of the spin polarization in the scanning photocurrent measurements follows $P \sim P_0 e^{-x/l_s}$, where $P_0$, $x$ and $l_s$ denote the peak polarization, the distance between the optical excitation and the electrode-TMD junction, and the spin free path, respectively (supplementary information). The representative contour demonstrated in Figure 4c yields $P_0 = 0.15 \pm 0.02$ and $l_s = 1.7 \pm 0.2 \mu m$ for holes, and $P_0 = 0.07 \pm 0.01$ and $l_s = 1.3 \pm 0.1 \mu m$ for electrons, respectively. The peak polarization $P_0$ is attributed to the spin polarization of the photo-carriers and the anisotropic magnetization resistance of the FM electrodes superimposed by the efficiency of the spin injection junction. If we assume the spin polarization of electrons at Fermi level of cobalt electrodes at 0.4 and the (up-bound) efficiency of the spin injection at 0.7[34], the near unit spin polarization of the photocurrent is estimated, surpassing all demonstrated in conventional semiconductors.

The micron size spin free path of electrons also implies a sizable spin-splitting in the conductance band edge which was theoretically predicted to be around 30meV.[26,35] The similar spin free paths of electrons ($1.3 \mu m$) and holes ($1.7 \mu m$) could be interpreted as the result of the close effective masses of electrons and holes and the large spin-splitting gaps at the conduction and valence band edges with respect to the thermal energy (10K~0.86meV) and the Fermi energy (around zero at intrinsic state) at the experimental conditions. Meanwhile figure 3e-f show the degree of spin polarization of holes is significantly higher than that of electrons, 15% vs 7%. This is consistent with the calculations that the spin splitting carries the same sign between conduction band and valence band monolayer WS$_2$ as shown in figure 1a.[26,27] As spin is conserved in the optical interband transition, electrons are pumped to the spin-split upper subband under near resonant excitations. The electron relaxation could take place through two channels, intravalley scattering where spin-flip is required or intervalley scattering where spin is conserved. Figure 3c-f show that the photocurrent difference $I_{\sigma+} - I_{\sigma-}$ and the degree of spin polarization carry the same polarity at both source and drain electrodes under the same FM magnetization. It implies that the spin-conserved intervalley scattering predominates the electron relaxation process and this also explains why the spin polarization of electrons (at source side) is weaker than that of holes (at drain side).

We also could estimate the magnitude of the spin lifetime from the spin free path. Given that the effective bias added on the channel is at the order of $V_{ds} - (E_g + E_{exciton\ binding}) \sim 2.5 eV$ where we assume the band bending at both contacts is roughly of the electronic band gap at most and the mobility of $0.1\text{-}1 cm^2 \cdot v/s$ of the devices (supplementary information), the spin free path indicates the estimated spin lifetime around $10^1 \sim 10^2$ nanosecond. This estimate is significantly larger than the valley lifetime estimated from PL polarization and pump-probe spectroscopy in which the valley lifetime of excitons instead of free carriers is probed.[36,37] Note that the electron-hole exchange interaction provides the major channel for excitons' spin/valley depolarization.[38] Whereas, the exchange interactions are greatly suppressed in oppositely drifting free carriers in a nearly intrinsic state, and consequently the free carriers presumably show significantly longer spin/valley lifetime.

In summary, we have demonstrated the fully spin-polarized photocurrents in monolayer $WS_2$ by controlling the polarization field of optical excitations. The spin polarization is well understood as the result of valley-dependent optical selection rules and spin-valley locking in monolayer TMDs. The demonstrated micron size spin free path and spin lifetime in range of $10^1 \sim 10^2$ nanosecond, and the unique spin-valley locking make monolayer TMDs a promising candidate for spintronics applications.

**Method:**

Monolayer $WS_2$ was mechanically exfoliated from an unintentionally doped bulk single crystal onto a silicon substrate with 300nm thick silicon oxide. A 5μm channel FET structure was fabricated with standard optical lithography. To make the tunneling barrier of ultrathin $Al_2O_3$ film, 1nm thick aluminum film was deposited with an effusion cell at base pressure of $2 \times 10^{-6}$ Pa and followed by oxidation in pure oxygen at 1bar for 24 hours. Ferromagnetic electrodes are made of a thin film superlattice of 20 alternating Co (4.5Å)/Pd(15Å) layers deposited by miniature e-beam evaporators with deposition rate of 1Å/min(Co) and 0.25Å/min(Pd) in a metal molecular beam epitaxy (MBE) system. The magnetization of the electrodes was *in-situ* characterized at 10K with a polar MOKE setup where the laser beam of 633nm is normally incident on the electrodes. The magnetization of FM electrodes in photocurrent measurements was switched by an external out-of-plane magnetic field of 0.1T.

The electric transport characterization and photocurrent measurements were carried out in an optical cryostat (sample in vacuum). The photocurrent measurements were conducted with a FET structure under 'off' state $V_G = -80V$ in a magnetic field free environment at 10K. The photocurrent was generated at a source-drain bias $V_{DS} = -5V$ under the near-resonance excitation of 2.09eV. The source-drain current is feed to a preamplifier with input impedance of 100KΩ close to the sample side. The laser was focused through a 50X objective lens onto a spot of 1μm and the excitation power is kept below 150μW. The photocurrents and the circular dichroism were monitored simultaneously with a photo-elastic modulator (50KHz) and two sets of lock-in amplifiers which extract both the photocurrents and the difference between two helicities. So the potential effects due to sample inhomogeneity were minimized.

The peak polarization and spin free path are extracted from fitting 12 sets of photocurrent polarization contours.

**Acknowledgement:** The work is supported by GRF (17300415), Area of excellency (AoE/P-04/08), CRF(HKU9/CRF/13G) of Hong Kong Research Grant Council and SRT on New Materials of The University of Hong Kong

**Figure**

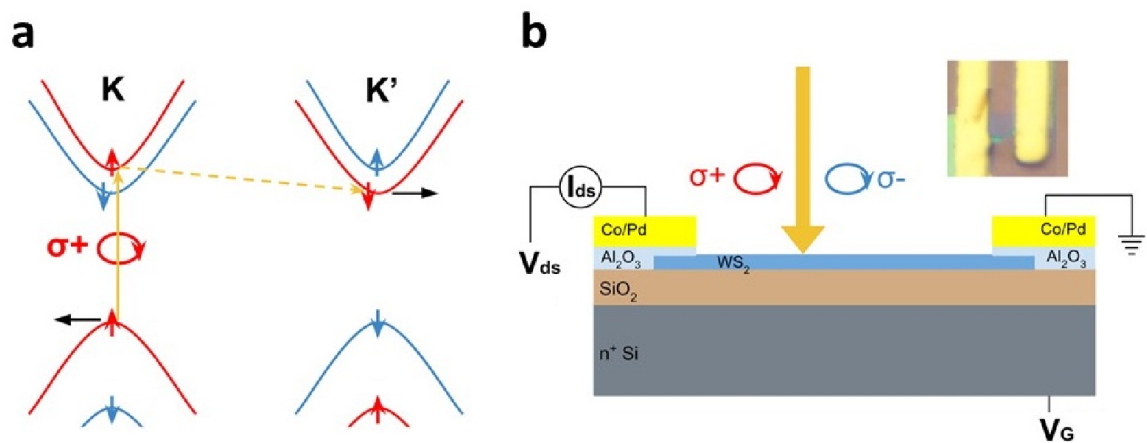

Figure 1. (a) Schematics of valley-dependent optical selection rules at k and k' valleys in the momentum space of monolayer TMDs and spin-valley locking, and the proposed mechanism of the spin-resolved photocurrent measurement with the ferromagnetic electrodes. The spin-splitting at conduction band (~0.03eV) and valence band (~0.45eV) are disproportionally sketched for clarity. (b) Schematic of monolayer $WS_2$ devices for spin-polarized photocurrent measurements. Inset is an optical image of the representative devices.

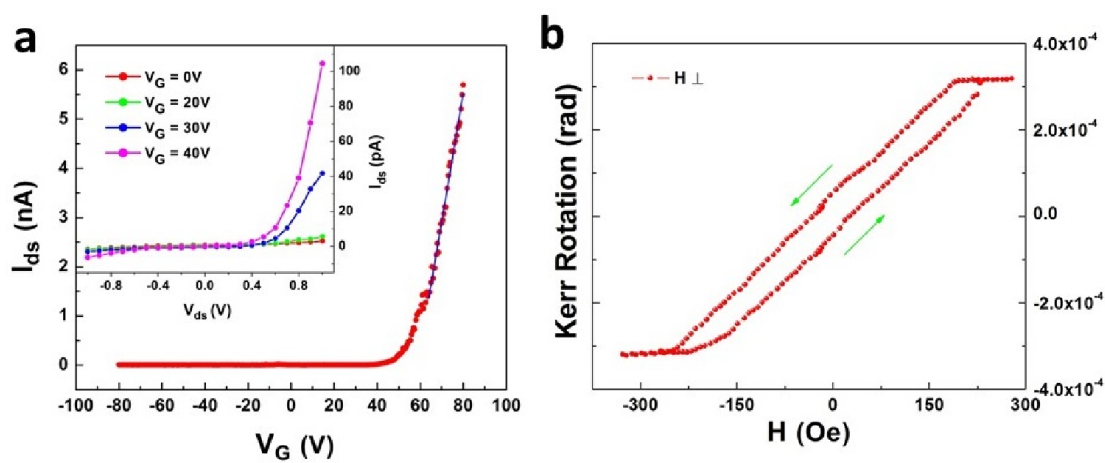

Figure 2. Transport characteristics (a) $I_{ds} - V_G$ curve and standard I-V characteristic of the device at 10K. (b) Polar MOKE measurement of Co/Pd layered FM electrodes at 10K with external magnetic field perpendicular to the sample surface. The magnetic hysteresis loop clearly shows a ferromagnetic behavior with perpendicular magnetization anisotropy (PMA).

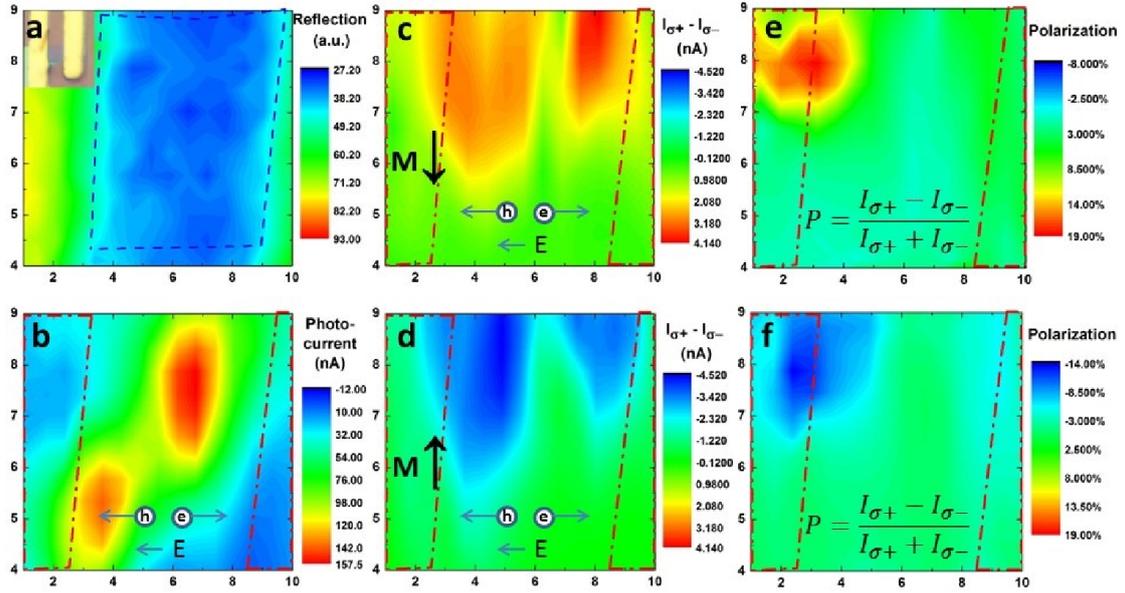

Figure 3. (a) Laser scanning reflection image of the photo-current device. The areas outlined with red dash line are FM electrodes. Inset is the corresponding optical image. (b) Photocurrent map with a scanning linearly polarized excitation under bias $V_G = -80V$ and $V_{DS} = -5V$. (c-d) The differential photocurrents $(I_{\sigma+} - I_{\sigma-})$ between the σ+ and σ- circularly polarized excitations through the FM electrodes with opposite magnetization M↑ and M↓ under zero magnetic field. The difference changes sign at opposite magnetizations. The photocurrent difference $I_{\sigma+} - I_{\sigma-}$ keeps the same polarity at both source and drain electrodes under the same FM magnetization. (e-f) The degree of the photocurrent polarization $P = (I_{\sigma+} - I_{\sigma-})/(I_{\sigma+} + I_{\sigma-})$ through the FM electrodes with opposite magnetization M↑ and M↓.

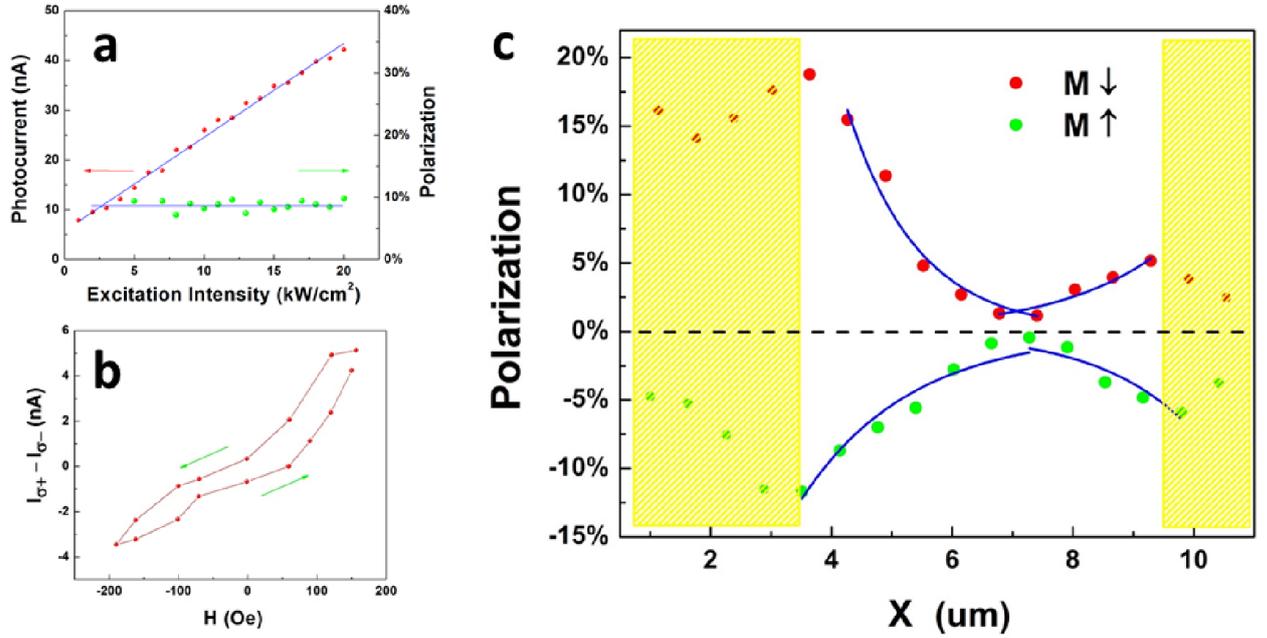

Figure 4. (a) The photocurrent and the degree of photocurrent polarization P as a function of the excitation intensity. (b) The photocurrent difference between circularly polarized excitations with opposite helicities as a function of external magnetic field along out-of-plane direction. The photocurrent difference shows a ferromagnetism like loop which is consistent with the magnetization of the FM electrodes. (c) Representative photocurrent polarization $P$ as a function of the distance from the FM electrodes with opposite magnetization M↑ and M↓. The hatch area labels the FM electrodes. The fit curve (blue) assuming $P \sim P_0 e^{-(x-x_0)/l_s}$ yields peak polarization $P_0 = 0.15 \pm 0.02$ and spin free path $l_s = 1.7 \pm 0.2 \mu m$ for holes, and $P_0 = 0.07 \pm 0.01$ and $l_s = 1.3 \pm 0.1 \mu m$ for electrons, respectively.